\documentclass[11pt]{article}
\usepackage{moriond,epsfig}

\long\def\Journal#1#2#3#4{{#1} {\bf #2}, #3 (#4)}

\def\NIMA{{\em Nucl.\,Instrum.\,Meth.\,}A}

\def\NPA{{\em Nucl.\,Phys.\,}A}
\def\PLB{{\em Phys.\,Lett.\,}B}
\def\PRL{\em Phys.\,Rev.\,Lett.\,}
\def\PRD{{\em Phys.\,Rev.\,}D}
\def\ZPC{{\em Z.\,Phys.\,}C}

\def\be{\begin{equation}}
\def\ee{\end{equation}}
\def\bea{\begin{eqnarray}}
\def\eea{\end{eqnarray}}

\def\bb{b\bar b}
\def\cc{c\bar c}
\def\BR{{\cal B}}
\def\etal{{\em et al.}}
\def\LL{l^+l^-}

\begin{document}
\vspace*{2cm}
\title{CLEO RESULTS ON TRANSITIONS IN HEAVY QUARKONIA
\footnote{
Presented at 40$^{th}$ Rencontres De Moriond 
On QCD And High Energy Hadronic Interactions,  
12-19 Mar 2005, La Thuile, Aosta Valley, Italy.}
}

\author{ T. SKWARNICKI }

\address{Department of Physics, 201 Physics Building, Syracuse University\\
Syracuse, NY 13244, USA}

\maketitle\abstracts{
Recent CLEO results on electromagnetic and hadronic transitions
in charmonium and bottomonium systems are reviewed.
}

\section{Introduction}

\begin{figure}
\vskip-0.2cm
\psfig{figure=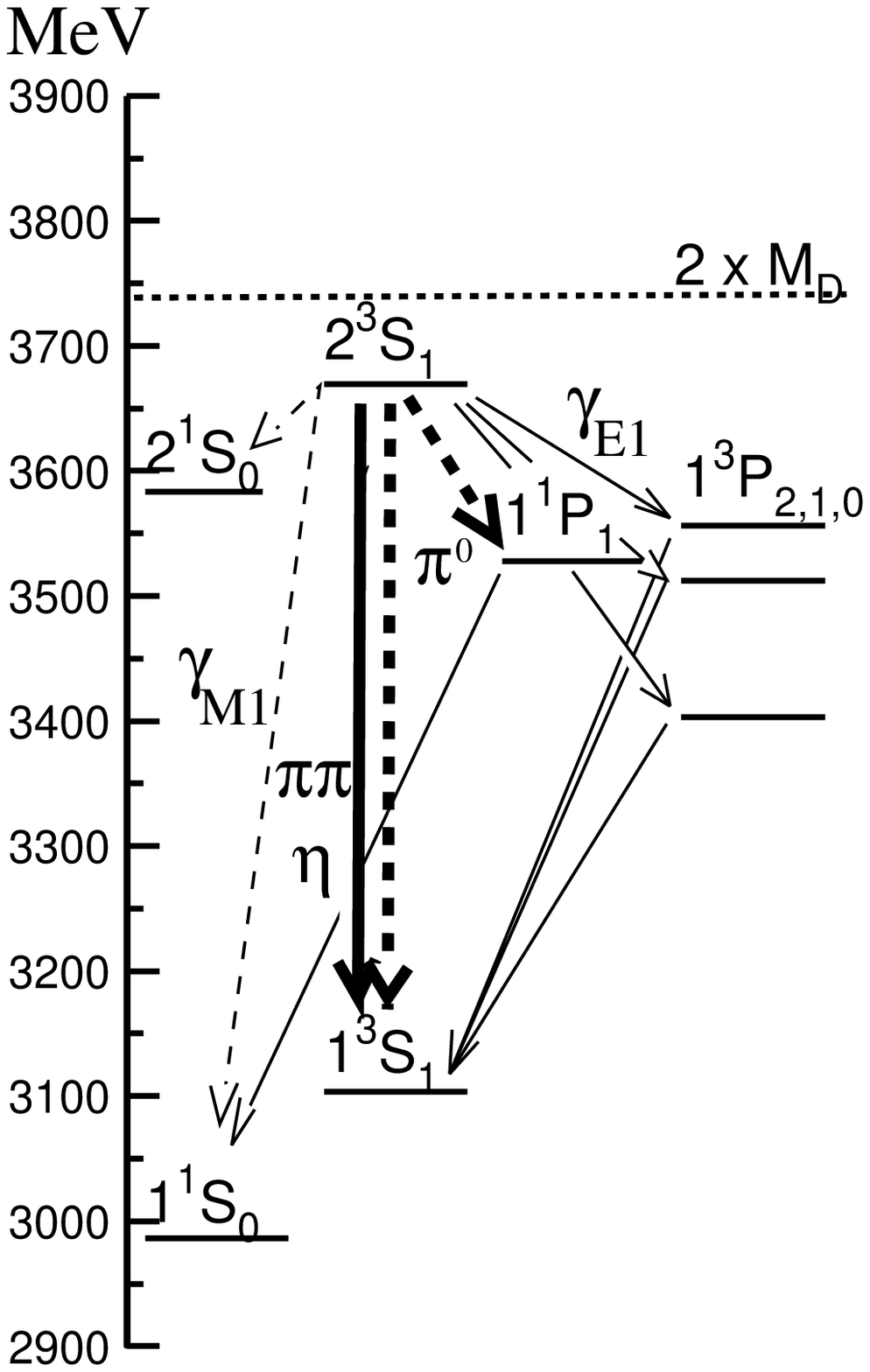,height=3.0in}
\psfig{figure=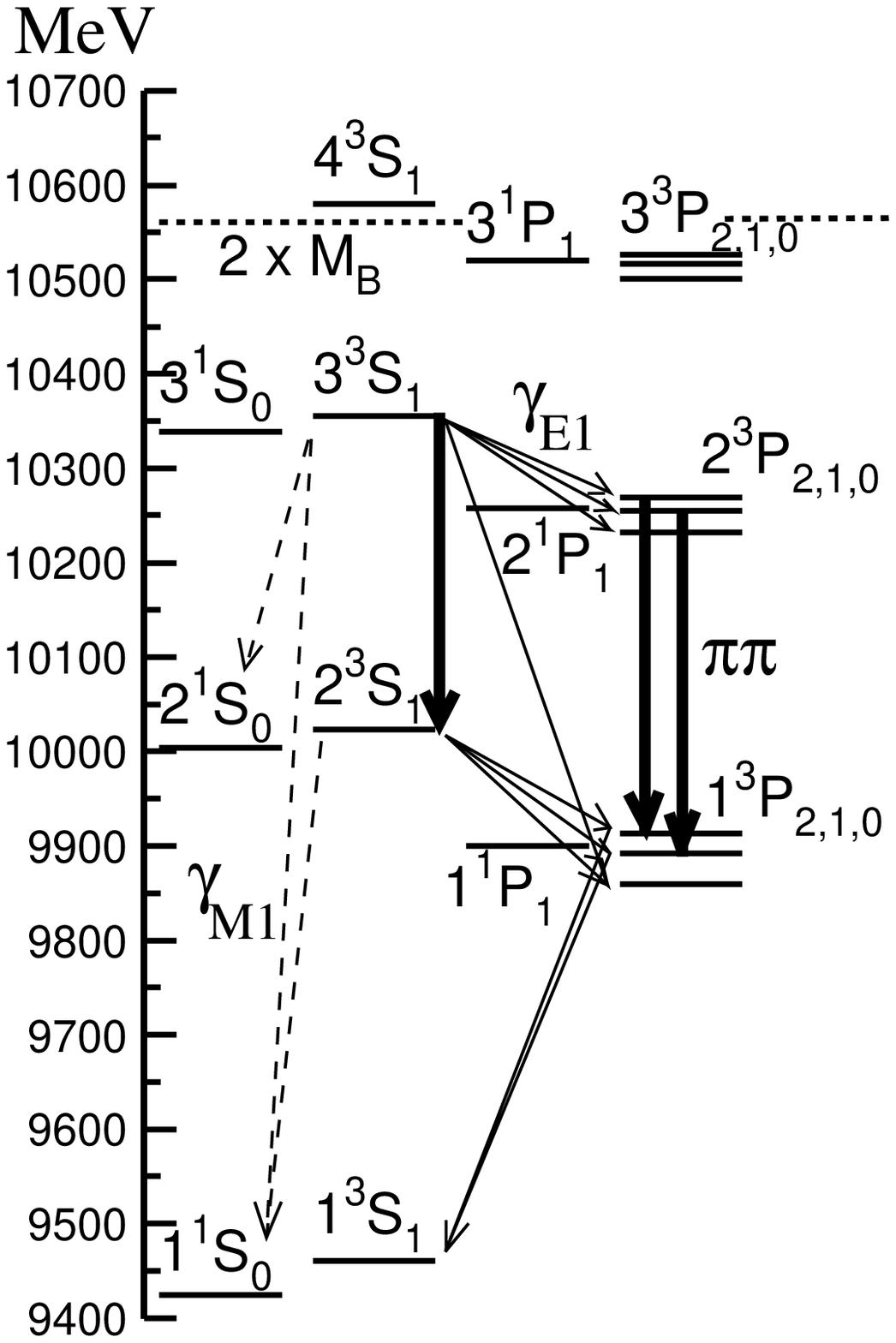,height=3.0in}
\vskip-1.0cm
\caption{Various transitions in: (a) $\cc$ (left),
and (b) $\bb$ (right) systems discussed
in this article. The E1 (M1) photon transitions are
indicated by the thin solid (dashed) lines.
The $\pi\pi$,$\eta$ ($\pi^0$) transitions are indicated
by the thick solid (dashed) lines.}
\label{fig:levels}
\vskip-0.2cm
\end{figure}

Heavy quarkonia~\cite{YellowReport} 
states ($\bb$ and $\cc$) below the open flavor threshold
live long enough that their
excitation level can be changed by emission of a photon or soft gluons
turning into light hadrons.
The triplet-$S$ states ($n^3S_1$) can be directly formed in $e^+e^-$ 
annihilation at the electron-positron storage rings. 
Then the other excitations levels can be observed via one or more
transitions (Fig.~\ref{fig:levels}).
The CLEO-III experiment~\cite{CLEOdetector,CLEORICH} 
collected large $\Upsilon(1S,2S,3S)$ samples 
at the end of CESR operations at the $\bb$ threshold region
($29$, $9$ and $6$ million resonant decays respectively).
These data samples are about a factor of 10 larger than previously
available. Then the CESR beam energy was lowered to the $\cc$ threshold
region. Three million $\psi(2S)$ resonant decays were recorded,
split about equally between the CLEO-III and CLEO-c detectors (the latter
has a small wire chamber replacing the CLEO-III silicon vertex detector).
Even though this is not the world's largest sample, it is nevertheless
unique, since CLEO is the first detector studying the charmonium
system with excellent detection of both charged particles and photons.
Excellent particle identification capabilities~\cite{CLEORICH}  
of the CLEO detector
are also important for some results presented here.

\section{Observation of $h_c(1^1P_1)$ State}

Spin-spin forces in heavy quarkonia are predicted to be short-range. 
Thus, while significant hyperfine splitting is observed for charmonium
$S$-states (e.g., 116 MeV for $n=1$), the mass splitting between the singlet  
state ($h_c(1^1P_1)$) and the center-of-gravity
of the spin-triplet states ($\sum_J (2J+1)\,m(\chi_{c}(1^3P_J))/\sum_J (2J+1)$)
is expected to be small.  
The $h_c(1^1P_1)$ was sighted previously twice
in $\bar pp$ annihilation at two different masses~\cite{hcold}
with marginal statistical significance. Higher statistics searches
disproved these observations.
We present highly significant evidence for this state, settling the
question about its mass.

\begin{figure}[htbp]
\vskip-0.2cm
\psfig{figure=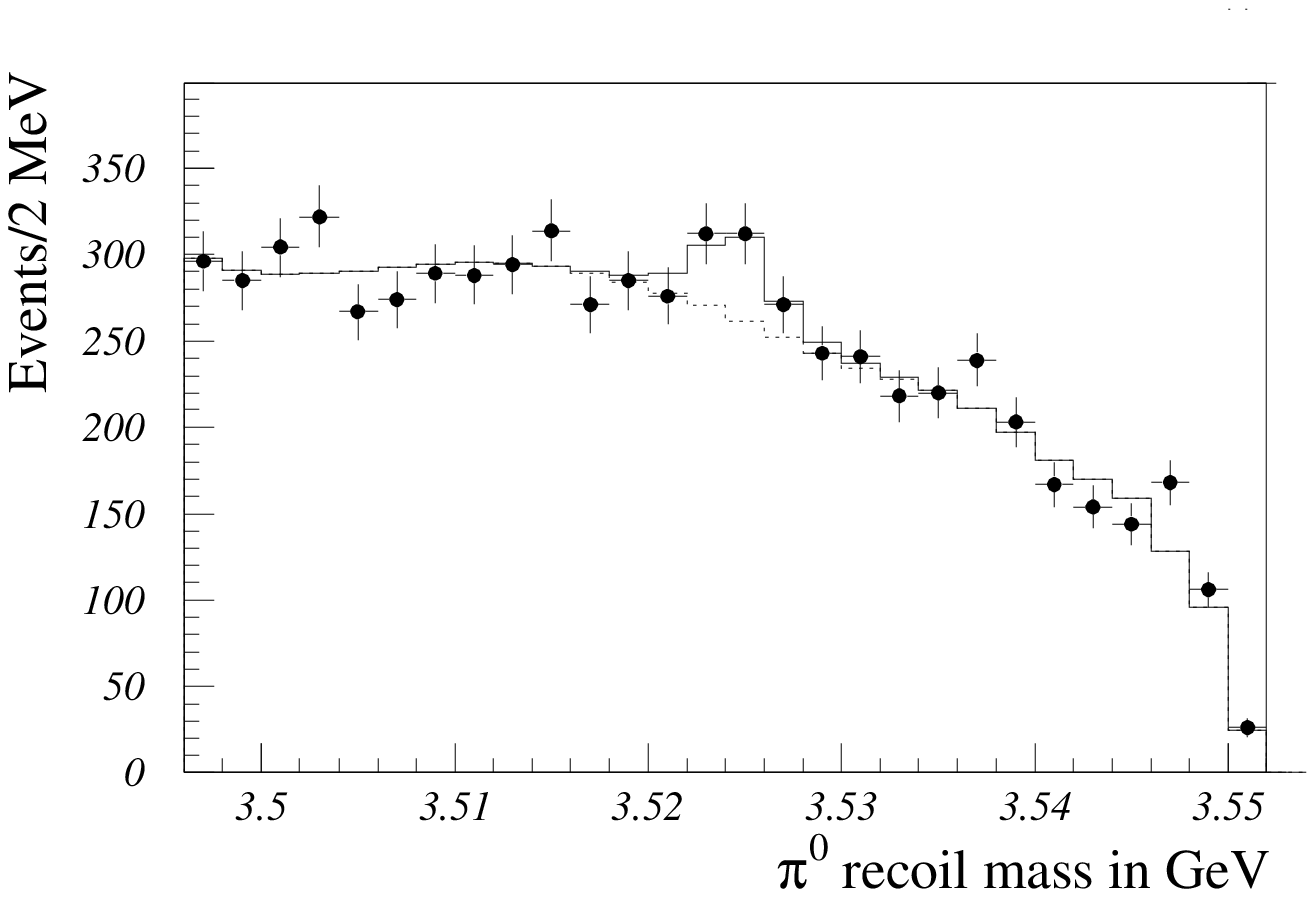,width=3.5in} \hskip-0.5cm
\psfig{figure=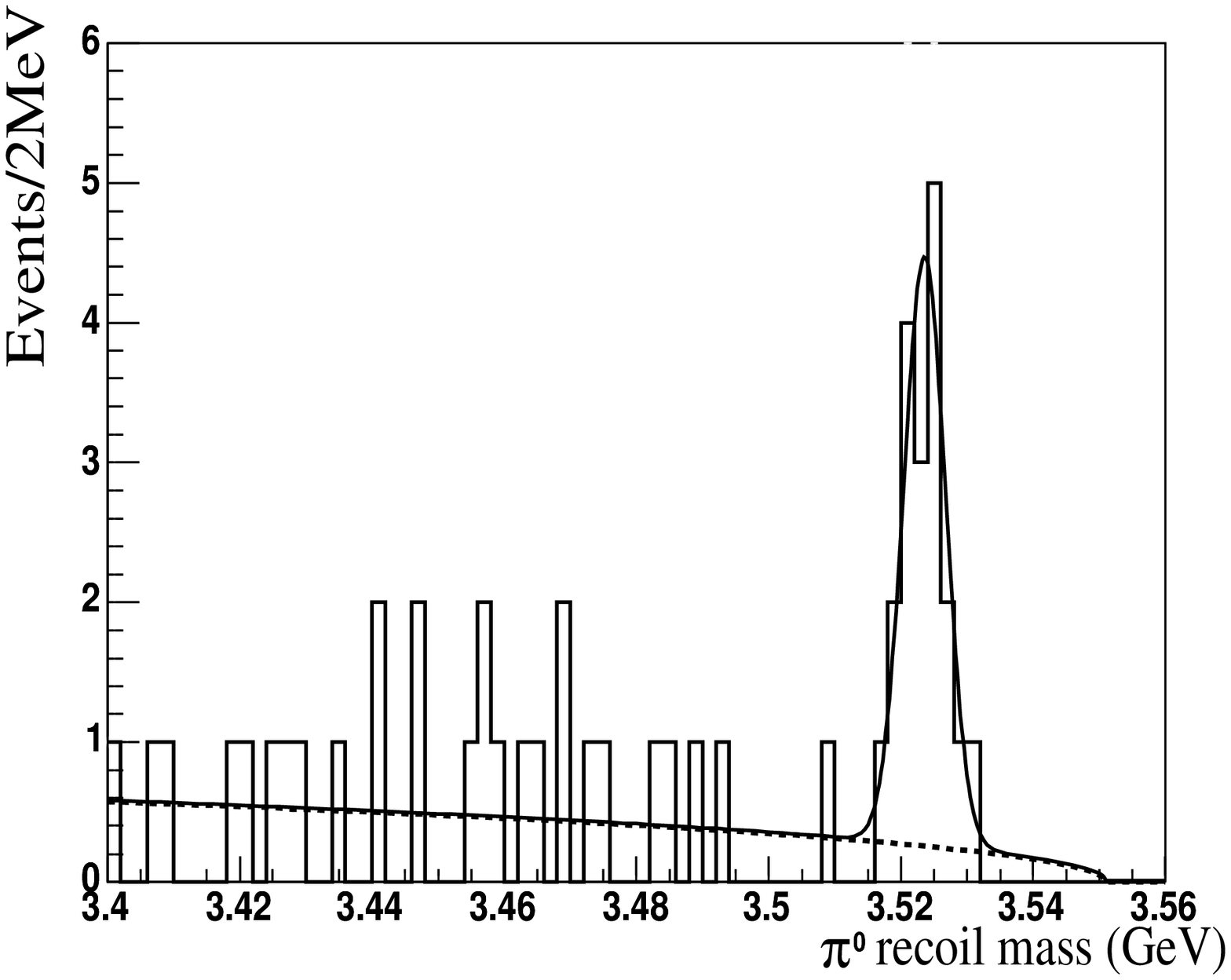,width=3.0in}
\vskip-0.4cm
\caption{Recoil mass against the reconstructed $\pi^0$ in:
(a) inclusive (left), and (b) exclusive (right)
search for the $h_c$
state. The fits are superimposed on the data.}
\label{fig:hc}
\vskip-0.2cm
\end{figure}

We have observed the $h_c(1P)$ state in isospin violating $\pi^0$ transitions
from the $\psi(2S)$ resonance, followed by a highly favored E1 photon
transition, $h_c(1P)\to\gamma\eta_c(1S)$ (see Fig.~\ref{fig:levels}a).  
Two essentially statistically independent approaches are used.
In the inclusive approach, the $\eta_c(1S)$ is
allowed to decay to anything. This approach results in a higher signal 
efficiency but also higher backgrounds.
After imposing consistency of the reconstructed
$\pi^0(\to\gamma\gamma)\gamma$ pair
with the $\psi(2S)$ to $\eta_c(1S)$ transition, 
the $\pi^0$-recoil mass is plotted (Fig.~\ref{fig:hc}a). The photon
four-vectors in the $\pi^0$ decay are constrained to the $\pi^0$ mass,
substantially improving the recoil mass resolution.
A peak of $150\pm40$ events, with a significance of 3.8 standard deviations,
is observed. 

In the second, exclusive, approach the $\eta_c(1S)$ is reconstructed
in one of the following decay modes: 
$K^0_S K^{\pm}\pi^{\mp}$, 
$K^0_L K^{\pm}\pi^{\mp}$, 
$K^+K^-\pi^+\pi^-$, 
$\pi^+\pi^-\pi^+\pi^-$, 
$K^+K^-\pi^0$, 
$K^+K^-\eta(\to\gamma\gamma\,\,{\rm or}\,\to\pi^+\pi^-\pi^0)$.
Particle ID capabilities of the CLEO detector (RICH~\cite{CLEORICH} 
and dE/dX) are critical in this analysis.
The $\eta_c(1S)$ reconstruction was optimized on the
hindered M1 photon transitions: $\psi(2S)\to\gamma\eta_c(1S)$ 
(see Fig.~\ref{fig:levels}a).
This approach results in excellent background suppression, but also in
smaller signal efficiency.
The $\pi^0-$recoil mass for the exclusive analysis is plotted in
Fig.~\ref{fig:hc}b.
A peak of $17.5\pm4.5$ events is observed at the mass
consistent with the inclusive analysis.
The probability of the background fluctuating up to produce this peak
is equivalent to a signal significance of 6.1 standard deviations.

The average of the inclusive and exclusive mass measurements,
$3524.4\pm0.6\pm0.4$ MeV, 
is $1.0\pm0.6\pm0.4$ MeV below the center-of-gravity of the
$\chi_{cJ}(1P)$ states, confirming the conventional
picture of spin-spin interactions. 
The measured product branching ratio, 
$\BR(\psi(2S)\to\pi^0 h_c(1P))\times
 \BR(h_c(1P)\to\gamma\eta_c(1S))=$ $(4.0\pm0.8\pm0.7)\times10^{-4}$,
is in the midrange of the theoretical predictions~\cite{hctheory}, 
which vary by 2 orders of magnitude due to 
difficulties in predicting the $\pi^0$ transition width.

\section{Survey of $\psi(2S)$ to $J/\psi(1S)$ Transitions}

We have performed a survey of $\psi(2S)$ to $J/\psi(1S)$ transitions, 
tagging $J/\psi$ by its annihilation to electron or muon pairs ($\LL$).
The $\BR(\psi(2S)\to X J/\psi(1S))$ is measured 
from the $J/\psi$ peak observed in the inclusive di-lepton mass distribution. 
Transition branching ratios for individual channels are
measured by full reconstruction of the following exclusive event
samples: $\pi^+\pi^-\LL$, $\pi^0\pi^0\LL$,  
$\eta(\to\gamma\gamma\,\,{\rm or}\,\to\pi^+\pi^-\pi^0)\LL$,
$\pi^0(\to\gamma\gamma)\LL$,
$\gamma\chi_{cJ}\to\gamma\gamma\LL$ (see Fig.~\ref{fig:levels}a).
The backgrounds are small and dominated by feed-across between the
transition modes. They are subtracted using Monte 
Carlo simulations.
The large statistics, the small backgrounds and the large,
well-understood detector acceptance result in the precision 
measurements. The results are
compared to previous measurements in Table~\ref{tab:psiptojpsix}.
A more detailed description of this analysis can be found
elsewhere~\cite{cleopsiptojpsi}.
These are the most precise
measurements to date. The difference between the inclusive
and the sum over exclusive branching ratios is $(0.6\pm0.4)\%$,
leaving little room for other, yet undetected modes.
Unlike previous measurements, the $\pi^0\pi^0$ rate is
half of the $\pi^+\pi^-$ rate, as expected from the isospin symmetry.
The branching ratios for two-photon cascades via the $\chi_{c0,1}$
states are significantly higher than previously measured, which 
leads to significantly larger rates for 
$\chi_{c0,1}\to\gamma J/\psi$.

\begin{table}[htp]
\newcommand{\gev}{\,\mbox{GeV}}
\newcommand{\mev}{\,\mbox{MeV}}
\newcommand{\invpb}{\,\mbox{pb}^{-1}}
\newcommand{\dilep}{\ell^+\ell^-}
\newcommand{\diel}{e^+e^-}
\newcommand{\dimu}{\mu^+\mu^-}
\newcommand{\jpsi}{J/\psi}
\newcommand{\pp}{\psi(2S)}
\newcommand{\chicJ}{\chi_{cJ}}
\newcommand{\egammalow}{E_{\gamma\mbox{-low}}}
\newcommand{\jpsiplusany}{X\jpsi}
\newcommand{\PiPiJ}{$\pi^+\pi^-\jpsi$}
\newcommand{\PizPizJ}{$\pi^0\pi^0\jpsi$}
\newcommand{\PizJ}{$\pi^0\jpsi$}
\newcommand{\EtaJ}{$\eta\jpsi$}
\newcommand{\EtaJGG}{$\eta(\to \gamma\gamma)\jpsi$}
\newcommand{\EtaJThreePi}{$\eta(\to \pi^+\pi^-\pi^0)\jpsi$}
\newcommand{\XJ}{$X\jpsi$}
\newcommand{\OnePiJ}{$\pi^\pm(\pi^\mp)\jpsi$}
\newcommand{\OnePizJ}{$\pi^0(\pi^0)\jpsi$}
\newcommand{\ChicZero}{$\gamma \chi_{c0} \to \gamma\gamma\jpsi$}
\newcommand{\ChicOne}{$\gamma \chi_{c1} \to \gamma\gamma\jpsi$}
\newcommand{\ChicTwo}{$\gamma \chi_{c2} \to \gamma\gamma\jpsi$}
\newcommand{\chicZero}{$\BR(\chi_{c0} \to \gamma\jpsi)$}
\newcommand{\chicOne}{$\BR(\chi_{c1} \to \gamma\jpsi)$}
\newcommand{\chicTwo}{$\BR(\chi_{c2} \to \gamma\jpsi)$}
\newcommand{\PiPiJRhoPi}{$\pi^+\pi^-\jpsi,\ \jpsi \to \rho\pi $}
\newcommand{\PiPiJPiPi}{$\pi^+\pi^-\jpsi,\ \jpsi \to \pi\pi $}
\setlength{\tabcolsep}{0.4pc}
\catcode`?=\active \def?{\kern\digitwidth}
\begin{center}
\footnotesize
\def\1#1{\multicolumn{1}{c|}{#1}}
\begin{tabular}{|l|r|r|r|r|r|} 
 \hline 
 Channel & 
\multicolumn{3}{c|}{${\cal B}$ (\%) } 
& \multicolumn{2}{c|}{${\cal B}/{\cal B}_{\pi^+\pi^-\jpsi}$ (\%) } \\  
\cline{2-6}
   &  \1{CLEO} & \1{PDG 2004} & \1{E835} & \1{CLEO} & \1{BES} \\
 \hline 
          \PiPiJ
 & $ 33.54 \pm 0.14 \pm 1.10 $ & $31.7\pm1.1$ & $29.2\pm0.5\pm1.8$
 & & 
  \\ 
 \hline 
        \PizPizJ
 & $ 16.52 \pm 0.14 \pm 0.58 $ & $18.8\pm1.2$ & $16.7\pm0.5\pm1.4$ 
 & $ 49.24 \pm 0.47 \pm 0.86 $ & $57.0\pm0.9\pm0.3$ 
  \\ 
 \hline 
     $\eta\jpsi$
 & $  3.25 \pm 0.06 \pm 0.11 $ & $3.16\pm0.22$ & $2.8\pm0.2\pm0.2$
 & $  9.68 \pm 0.19 \pm 0.13 $ & $9.8\pm0.5\pm1.0$
  \\ 
 \hline 
           \PizJ
 & $  0.13 \pm 0.01 \pm 0.01 $ & $0.10\pm0.02$ &
 & $  0.39 \pm 0.04 \pm 0.01 $ & 
  \\ 
 \hline 
       \ChicZero
 & $  0.18 \pm 0.01 \pm 0.02 $ & $0.10\pm0.01$  &
 & $  0.55 \pm 0.04 \pm 0.06 $ & 
  \\ 
       \chicZero
 & \1{$  2.0 \pm 0.3$} & \1{$1.2\pm0.1$}  &   &   &
  \\ 
 \hline 
        \ChicOne
 & $  3.44 \pm 0.06 \pm 0.13 $ & $2.67\pm0.15$ & 
 & $ 10.24 \pm 0.17 \pm 0.23 $ & $12.6\pm0.3\pm3.8$ 
  \\ 
        \chicOne
 & \1{$  37.9 \pm 2.2 $} & \1{$31.6\pm3.3$} & & & 
  \\ 
 \hline 
        \ChicTwo
 & $  1.85 \pm 0.04 \pm 0.07 $ & $1.30\pm0.08$ & 
 & $  5.52 \pm 0.13 \pm 0.13 $ & $6.0\pm0.1\pm2.8$
  \\ 
        \chicTwo
 & \1{$  19.9 \pm 1.3 $} & \1{$20.2\pm1.7$} & & &
  \\ 
 \hline 
             \XJ
 & $ 59.50 \pm 0.15 \pm 1.90 $ & $57.6\pm2.0$ & 
 & $1.77\pm0.01\pm0.02$  & $1.87\pm0.03\pm0.06$ 
  \\ 
 \hline 

\end{tabular} 
\end{center}
\caption{
The CLEO results~\protect\cite{cleopsiptojpsi}
for  $\psi(2S)$ to $J/\psi(1S)$ transitions
compared to the PDG fit values~\protect\cite{PDG}
and two recently published measurements by
BES~\protect\cite{BESpsiptojpsi} and E835~\protect\cite{E835psiptojpsi}.
\label{tab:psiptojpsix}
}
\end{table}

\section{First Evidence for $\chi_b(2P)\to\pi^+\pi^-\chi_b(1P)$
         Transitions}

\begin{figure}
\long\def\makemycaption#1#2{
   \vskip 10pt 
       \leftskip 0pt plus 1fil 
       \rightskip 0pt plus -1fil 
       \parfillskip 0pt plus 2fil 
       \footnotesize #1: #2\par
}
\parbox[b]{3.0in}{
\begin{center}
\begin{tabular}{|c|c|c|}
\hline
Region & $N_{\mathrm{bgd}}$ & $N_{\mathrm{obs}}$ \\
\hline
       & \multicolumn{2}{c|}{di-pion sample} \\
\hline
other-backgrounds & $37.6\pm5.8$ & 36 \\ 
$\Upsilon(2S)_\mathrm{bgd}$ & $7.6\pm1.9$ & 10 \\ 
signal	& $1.2\pm0.2$ & 7 \\ 
\hline
     & \multicolumn{2}{c|}{single-pion sample} \\
\hline
other-backgrounds & $11.3\pm2.6$ & 13 \\ 
$\Upsilon(2S)_\mathrm{bgd}$ & $19.4\pm4.9$ & 26 \\
signal & $3.3\pm0.7$ & 17 \\
\hline
\end{tabular}
\addtocounter{table}{1}
\end{center}
\makemycaption{Table 2.}{
The results for $\chi_b(2P)\to\pi^+\pi^-\chi_b(1P)$.
Number of the observed ($N_{\mathrm{obs}}$)
and estimated background events ($N_{\mathrm{bgd}}$)
are given in the other-backgrounds, $\Upsilon(2S)_\mathrm{bgd}$ and
signal regions (see the text).}
}
\qquad
\parbox[b]{2.5in}{
\begin{center} 
\psfig{figure=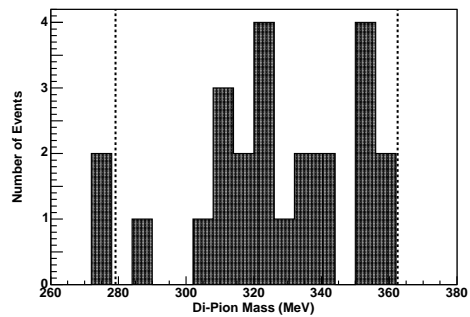,height=1.8in}
\quad
\end{center}
\vskip-0.4cm
\caption{Distribution of $M(\pi^+\pi^-)$ for selected 
$\chi_b(2P)\to\pi^+\pi^-\chi_b(1P)$ events (di-pion and single-pion
samples together). The dashed lines show the
kinematic limits. The estimated background level is $4.5\pm0.7$ events
($19\%$).}
\label{fig:pipi}
}
\end{figure}

The $\pi^+\pi^-$ transitions have been previously observed 
in the $\bb$ system between the $n^3S_1$ states ($n=1,2,3$).
Such transitions are also expected among the $n^3P_J$ states.
We are presenting the first evidence for these transitions.
The $\chi_b(2^3P_{J_2})$ states are produced by the E1
photon transitions from the $\Upsilon(3S)$ resonance 
(see Fig.~\ref{fig:levels}b).
The  $\chi_b(1^3P_{J_1})$ states are recognized by E1 photon transition
to the $\Upsilon(1S)$, followed by annihilation to $\LL$.
We look for the $\pi^+\pi^-$ transitions 
for $J_2=J_1=1$ or $2$, which are expected to
have the largest rate. 
The dominant backgrounds come from the other
transitions in the $\bb$ system, 
$\Upsilon(3S)\to\pi^+\pi^-\Upsilon(2S)$, 
$\Upsilon(2S)\to\gamma \chi_{bJ}(1P)$, 
$\chi_{bJ}(1P)\to\gamma\Upsilon(1S)$ 
(hereafter $\Upsilon(2S)_\mathrm{bgd}$) in particular (see Fig.~\ref{fig:levels}b).
The fully reconstructed $\gamma\pi^+\pi^-\gamma\LL$ events provide good
background suppression but suffer from the small signal efficiency 
($\epsilon\sim 4.5\%$), since
the soft pions often curl-up and escape detection in the 
tracking system. Therefore, we have also selected events with only one
detected pion ($\epsilon\sim 8.7\%$). For the di-pion events we define the signal,
$\Upsilon(2S)_\mathrm{bgd}$ and other-backgrounds regions by
cuts on the energy of the lower-energy photon
(identifying the $\chi_{b}(2P_{1,2})$ states) 
and the $\gamma_{low}\pi^+\pi^-$ -- $\gamma_{low}$ 
recoil mass-difference 
(identifying the $\chi_{b}(1P_{1,2})$ states).
For the single-pion events the latter is replaced by a cut on
the missing-mass of the event (reflecting the mass of the undetected
pion). The observed event yields are compared to the estimated
background rates in Table~2. 
After the background estimates are tuned to describe the observed
background levels in the other-backgrounds sidebands, the observed
event yields in the $\Upsilon(2S)_\mathrm{bgd}$
regions are also well reproduced.
In contrast, the signal region contains an excess of events,
which corresponds to a statistical significance of
6 standard deviations.
The $\pi^+\pi^-$ mass distribution in the signal region is plotted
in Fig.~\ref{fig:pipi}. All results presented in this section are
preliminary.

\section{Photon transitions} 

We have analyzed
inclusive photon spectra in the $\psi(2S)$, $\Upsilon(2S)$ and $\Upsilon(3S)$
data  for monochromatic photons due to
E1 and M1 photon transitions (see Fig.~\ref{fig:levels}). 
The results have been published and can be found elsewhere~\cite{Incl}.

From the measurements of photon energies in the dominant E1 transitions,
$n^3S_1\to\gamma (\! n\!-\!1\!)^3P_{2,1,0}$, ratios of the
fine splittings in the triplet-$P$ states, $(M_2-M_1)/(M_1-M_0)$,
are determined with a high precision: $0.490\pm0.002\pm0.003$ ($1P$ $\cc$),
$0.57\pm0.01\pm0.01$ ($1P$ $\bb$) and
$0.58\pm0.01\pm0.01$ ($2P$ $\bb$).
Somewhat surprisingly, the latter two are essentially equal.

In the non-relativistic limit, the E1 matrix elements for these
transitions are $J$ independent. Thus, a ratio of the branching ratios
($\BR(^3S_1\to\gamma ^3P_J)$)
corrected for the phase-space factors 
($(2J+1) E_\gamma^3$) is expected
to be 1 for any combination of $J$ values.  
The results are summarized in Table~\ref{tab:bratios}.
While the $(J=2)/(J=1)$ ratios in the $\bb$ system 
reproduce this expectation, the rates to the $J=0$ state are lower.
Relativistic corrections were predicted to be, in fact, the largest for
the transitions to $^3P_0$ state~\cite{relcor}.
The ratios in the $\cc$ system are far from the non-relativistic
prediction, apparently affected by the lighter quark mass and the
$2^3S_1-1^3D_1$ mixing.

\begin{table}[htbp]
\begin{center}
\begin{tabular}{|c|c|c|c|} 
 \hline 
Final state & (J=2)/(J=1) & (J=0)/(J=1) & (J=0)/(J=2) \\	
\hline
$\chi_b(2P)$ & $1.00\pm0.01\pm0.05$ & $0.76\pm0.02\pm0.07$ &
$0.76\pm0.02\pm0.09$ \\
$\chi_b(1P)$ & $1.01\pm0.02\pm0.08$ & $0.82\pm0.02\pm0.06$ &
$0.81\pm0.02\pm0.11$ \\
$\chi_c(1P)$ & $1.50\pm0.02\pm0.05$ & $0.86\pm0.01\pm0.06$ &
$0.59\pm0.01\pm0.05$ \\
\hline
\end{tabular} 
\end{center}
\vskip-0.2cm
\caption{Ratio of 
$\BR(n^3S_1\to\gamma(\! n\!-\!1\!)^3P_{J})/ (2J+1) E_\gamma^3 $
as measured by CLEO
for various $J$ combinations in the charmonium and bottomonium
systems.}
\label{tab:bratios}
\vskip-0.2cm
\end{table}

\begin{figure}[hbtp]
\vskip-1.5cm
\psfig{figure=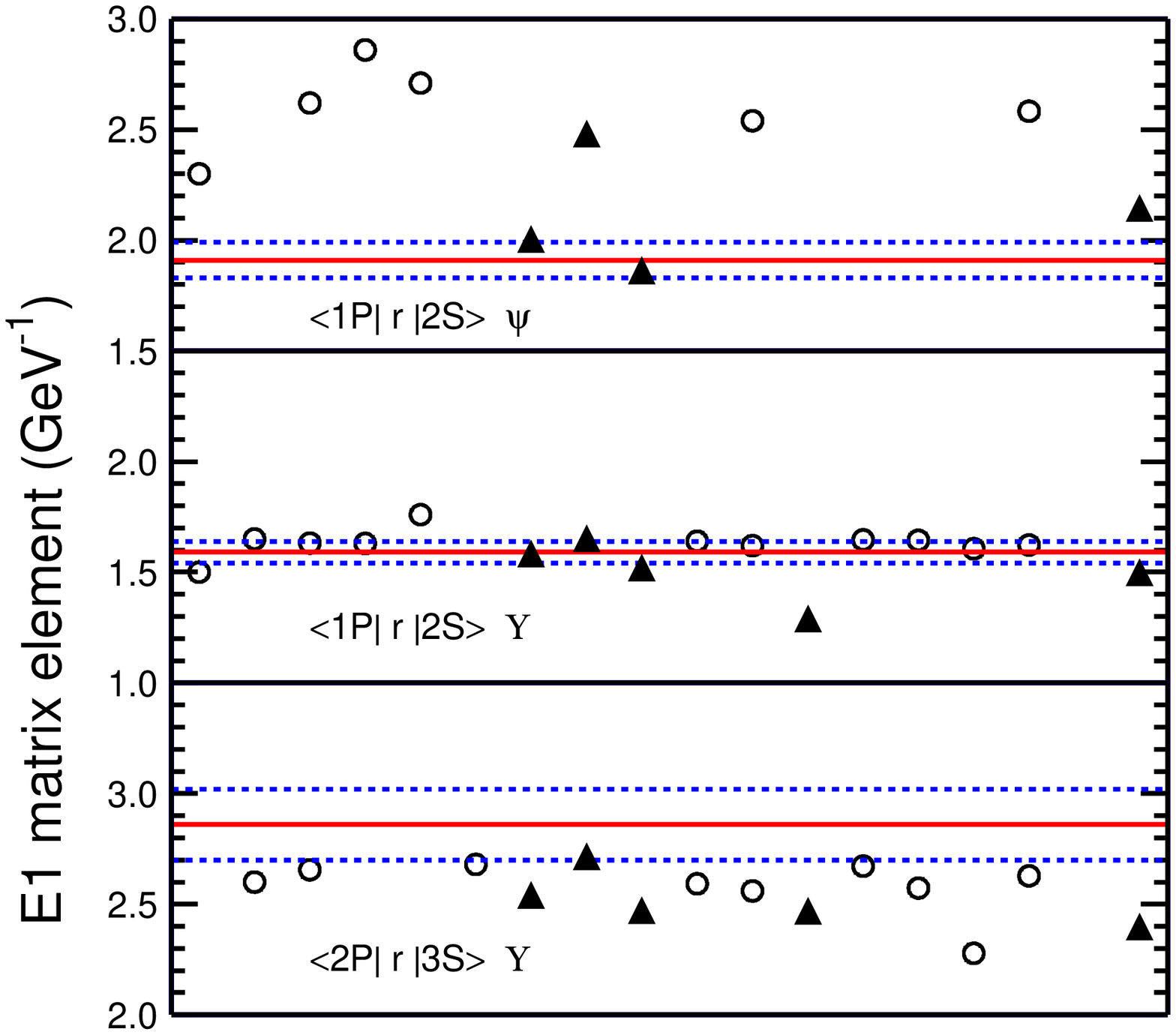,height=3in}
\hskip-0.4cm
\raisebox{-1.1cm}{\psfig{figure=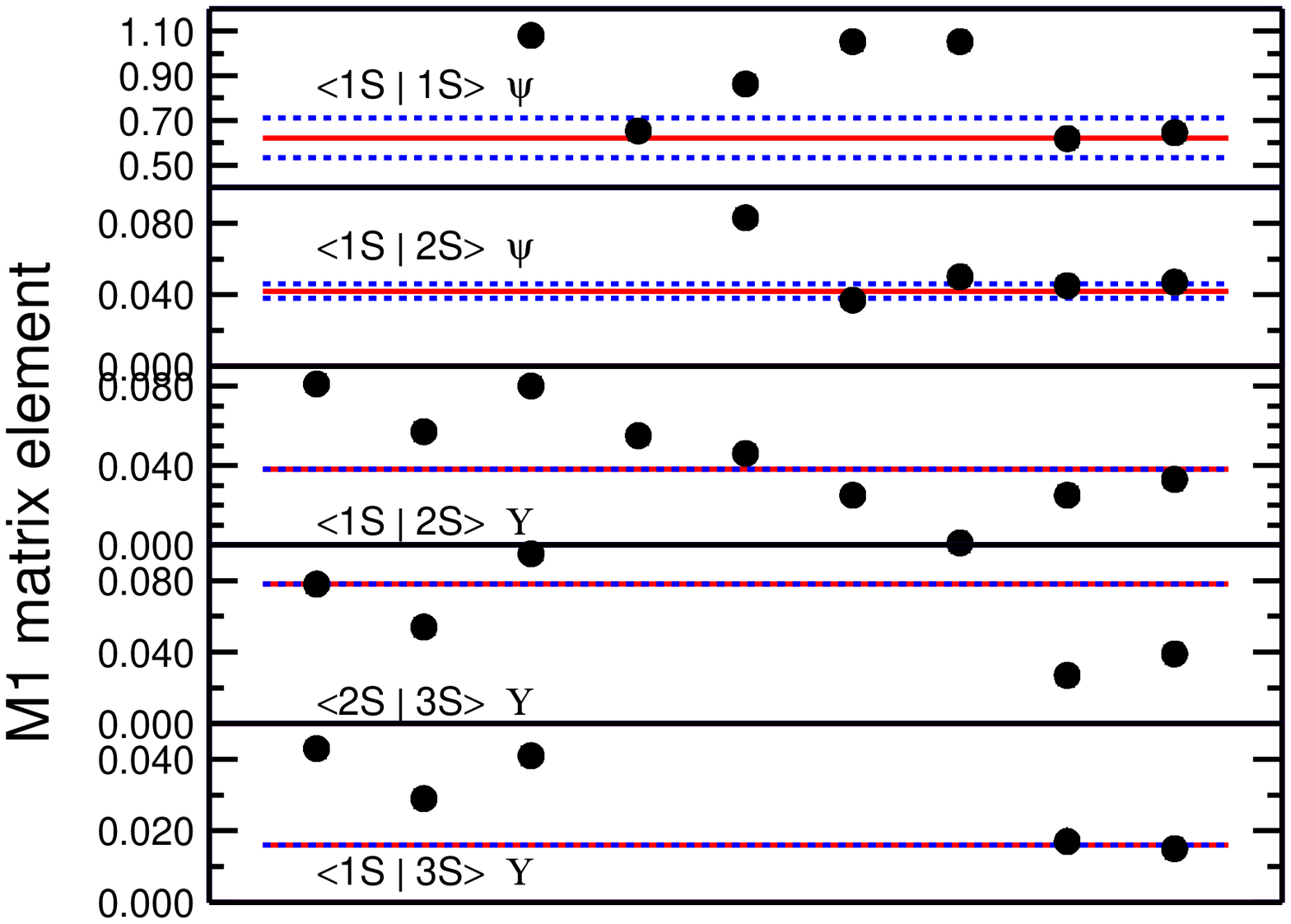,height=3.5in}}
\vskip-2.0cm
\caption{Measured and predicted values of matrix elements
         for: (a) E1 (left), and (b) M1 (right) transitions in heavy quarkonia.
         The E1 rates are  averaged over
         different spins of the triplet $P$ states. 
         The measured values are calculated
         from the CLEO branching ratio results (and
         total widths of the triplet $S$ states~\protect\cite{PDG,Bmm}), 
         except for the direct M1 transition, $2S\to 1S$, where the
         world average branching ratio~\protect\cite{PDG} is used.
         The central values and error bars for the measured values
         are indicated by the solid and dashed
         lines respectively. The solid lines for the M1 transitions
         in the $\Upsilon$ system show the experimental upper limits. 
         Circles (triangles) show non-relativistic 
         (relativistic) calculations. The relativistic calculations are 
         averaged over spins with the same weights as the data.
         The predictions~\protect\cite{e1m1preds} 
         are ordered according to the publication date.}
\label{fig:e1m1}
\end{figure}

The absolute values of the branching ratios are also significantly below
the non-relativistic predictions for the $\cc$ system. Relativistic
corrections are needed to explain the observed rates, 
as illustrated in Fig.~\ref{fig:e1m1}a.
In contrast, the relativistic correction in the $\bb$ system are 
not large and even non-relativistic calculations give a reasonable
description of the data. 
This is true only for the dominant
E1 transitions. The E1 matrix elements for the
$3^3S_1\to\gamma 1^3P_J$ transitions are expected to be small,
reflecting large cancellations in the integral of the dipole operator
between the $3^3S_1$ and $1^3P_J$ states. The relativistic corrections,
and therefore the $J$ dependence, are expected to be large.
We have measured the $J=0$ rate for the first time.
The theoretical predictions are scattered in a wide range and only
a few models match our data well~\cite{Incl}.

While we have confirmed the hindered M1 transition 
$\psi(2S)\to\gamma\eta_c(1S)$, previously observed 
by Crystal Ball~\cite{CBetac},
their signal for the direct M1 transition 
$\psi(2S)\to\gamma\eta_c(2S)$~\cite{CBetacp}
is not observed in our data. This is not surprising in view of
the recent $\eta_c(2S)$ mass measurements~\cite{etacp}, 
which are inconsistent
with the $\eta_c(2S)$ mass claimed by Crystal Ball.
Searches for hindered M1 transitions in the $\bb$ system resulted in
upper limits only, thus no singlet $\bb$ state has been observed to date.
Only the most recent theoretical estimates of the expected M1 rates are
consistent with all $\cc$ and $\bb$ data, and only marginally so with 
our limit on $\BR(\Upsilon(3S)\to\gamma\eta_b(1S))$
(see Fig.~\ref{fig:e1m1}b).

\section*{Acknowledgments}

The author thanks his CLEO colleagues for 
the input to this article.
This work was supported by 
the National Science Foundation and
the U.S. Department of Energy.


\section*{References}

\begin{thebibliography}{99}

\bibitem{YellowReport}
Quarkonium Working Group, N.~Brambilla \etal, hep-ph/0412158.

\bibitem{CLEOdetector}
CLEO Collaboration,
Y.~Kubota \etal,
\Journal{\NIMA}{320}{66}{1992};
G.~Viehhauser \etal,
\Journal{\NIMA}{462}{146}{2001};
D.~Peterson \etal, 
\Journal{\NIMA}{478}{142}{2002}.

\bibitem{CLEORICH}
M.~Artuso \etal,
\Journal{\NIMA}{502}{91}{2003}.

\bibitem{hcold}
R704 Collaboration, C.~Baglin \etal, 
\Journal{\PLB}{171}{135}{1986};
E760 Collaboration, T.~A.~Armstrong \etal, \Journal{\PRL}{69}{2337}{1992}.

\bibitem{hctheory}
Y.~P.~Kuang, \Journal{\PRD}{65}{094024}{2002};
S. Godfrey, J.L. Rosner, \Journal{\PRD}{66}{014012}{2002};
P.~Ko,  \Journal{\PRD}{52}{1710}{1995}.

\bibitem{cleopsiptojpsi}
CLEO Collaboration, N.~E.~Adam  \etal,
arXiv:hep-ex/0503028.

\bibitem{PDG}
Particle Data Group,
S.~Eidelman \etal, \Journal{\PLB}{592}{1}{2004}.

\bibitem{BESpsiptojpsi}
BES Collaboration,  M.~Ablikim \etal,
\Journal{\PRD}{70}{012003}{2004}. 

\bibitem{E835psiptojpsi}
E835 Collaboration, M.~Andreotti \etal,
\Journal{\PRD}{71}{032006}{2005}.

\bibitem{Incl}
CLEO Collaboration, M.~Artuso \etal,
\Journal{\PRL} {94}{032001}{2005};
S.~B.~Athar \etal,
\Journal{\PRD}{70}{112002}{2004}.

\bibitem{relcor}
P.~Moxhay, J.~L.~Rosner, \Journal{\PRD}{28}{1132}{1983};
R.~McClary, N.~Byers, \Journal{\PRD}{28}{1692}{1983}.

\bibitem{CBetac}
Crystal Ball Collaboration,
J.E.~Gaiser \etal, 
\Journal{\PRD}{34}{711}{1986}.

\bibitem{CBetacp}
Crystal Ball Collaboration,
C.~Edwards \etal,
\Journal{\PRL}{48}{70}{1982}.

\bibitem{etacp}
Belle Collaboration, S.K. Choi \etal, 
\Journal{\PRL}{89}{102001}{2002}; 
CLEO Collaboration, D.M. Asner \etal, 
\Journal{\PRL}{92}{142001}{2004};
BaBar Collaboration, B. Aubert \etal, 
\Journal{\PRL}{92}{142002}{2004}.

\bibitem{Bmm}
CLEO Collaboration,
G.S. Adams \etal,
\Journal{\PRL}{94}{012001}{2005}. 

\bibitem{e1m1preds}
The following predictions for the E1 matrix
elements are displayed in Fig.~\protect\ref{fig:e1m1}a:
D.~Pignon, C.~A.~Piketty, \Journal{\PLB}{74}{108}{1978};   
E.~Eichten, K.~Gottfried, T.~Kinoshita, K.~D.~Lane, T.~M.~Yan,
\Journal{\PRD}{21}{203}{1980};
W.~Buchmuller, G.~Grunberg, S.-H.~Tye
\Journal{\PRL}{45}{103}{1980}, \Journal{\PRD}{24}{132}{1981};
C.~Quigg, J.~L.~Rosner, \Journal{\PRD}{23}{2625}{1981} 
(2 entries: $\cc$, $\bb$ potential respectively);
J.~Baacke, Y.~Igarashi, G.~Kasperidus, \Journal{\ZPC}{13}{131}{1982};
R.~McClary, N.~Byers, \Journal{\PRD}{28}{1692}{1983};
P.~Moxhay, J.~L.~Rosner, \Journal{\PRD}{28}{1132}{1983};
H.~Grotch, D.~A.~Owen, K.~J.~Sebastian, \Journal{\PRD}{30}{1924}{1984};
S.~N.~Gupta, S.~F.~Radford, W.~W.~Repko, 
  \Journal{\PRD}{26}{3305}{1982}, 
  \Journal{\PRD}{30}{2424}{1984};
S.~N.~Gupta, S.~F.~Radford, W.~W.~Repko, \Journal{\PRD}{34}{201}{1986};
M.~Bander, D.~Silverman, B.~Klima, U.~Maor,
  \Journal{\PLB}{134}{258}{1984}, \Journal{\PRD}{29}{2038}{1984},
  \Journal{\PRD}{36}{3401}{1987};
W.~Kwong, J.~L.~Rosner, \Journal{\PRD}{38}{279}{1988};
L.~P.~Fulcher, \Journal{\PRD}{37}{1259}{1988};
S.~N.~Gupta, W.~W.~Repko, C.~J.~Suchyta III, \Journal{\PRD}{39}{974}{1989};
L.~P.~Fulcher, \Journal{\PRD}{42}{2337}{1990};
A.~K.~Grant, J.~L.~Rosener, E.~Rynes, \Journal{\PRD}{47}{1981}{1993};
T.~A.~Lahde, \Journal{\NPA}{714}{183}{2003};
M1 matrix elements in Fig.~\protect\ref{fig:e1m1}b:
V.~Zambetakis, N.Byers, \Journal{\PRD}{28}{2908}{1983}; 
H.~Grotch, D.~A.~Owen, K.~J.~Sebastian, \Journal{\PRD}{30}{1924}{1984}
(2 entries: scalar and vector confinement potential); 
S.~Godfrey, N.~Isgur, \Journal{\PRD}{32}{189}{1985}
(2 entries: based on quoted transition moments and wave functions,
 respectively);
X.~Zhang, K.~J.~Sebastian, H.~Grotch, \Journal{\PRD}{44}{1606}{1991}
(2 entries: scalar-vector and pure scalar confinement potential); 
D.~Ebert, R.~N.~Faustov, V.~O.~Galkin \Journal{\PRD}{67}{014027}{2003}; 
T.~A.~Lahde, \Journal{\NPA}{714}{183}{2003}.
Values of the $\bb$ M1 matrix elements 
are displayed for the photon energies and $b$ quark mass assumed in
S.~Godfrey, J.~L.~Rosner, \Journal{\PRD}{64}{074011}{2001},
Erratum-ibid.\ \Journal{}{65}{039901}{2002}.


\end{thebibliography}
\end{document}